\documentclass[prb, twocolumn, superscriptaddress, aps, a4paper]{revtex4}
\usepackage{amssymb}
\usepackage{amsmath}
\usepackage{graphicx}
\usepackage{epstopdf}
\usepackage{multirow}

\begin{document}

\title{Interlayer couplings and the coexistence of antiferromagnetic and d-wave pairing order in multilayer cuprates}
\author{Wei-Qiang Chen}
\affiliation{Department of Physics, South University of Science and Technology of China, Shenzhen, Guangdong, China}
\affiliation{Department of Physics, and Center of Theoretical and Computational Physics, The University of Hong Kong,
 Hong Kong, China}
\author{J. Y. Gan}
\affiliation{Institute of Physics, Chinese Academy of Sciences P.O.Box 603, Beijing 100190, China}
\author{T. M. Rice}
\affiliation{Institut f\"{u}r Theoretische Physik, ETH Z\"{u}rich,CH-8093 Z\"{u}rich, Switzerland}
\affiliation{Department of Physics, and Center of Theoretical and Computational Physics, The University of Hong Kong,
 Hong Kong, China}
\affiliation{Condensed Matter Physics and Materials Science Department,Brookhaven National Laboratory, Upton, NY 11973, USA}
\author{F. C. Zhang}
\affiliation{Department of Physics, and Center of Theoretical and Computational Physics, The University of Hong Kong,
 Hong Kong, China}

\begin{abstract}
   A more extended low density region of coexisting uniform antiferromagnetism and d-wave superconductivity has been reported in multilayer cuprates, when compared to single or bilayer cuprates. This coexistence could be due to the enhanced screening of random potential modulations in inner layers or to the interlayer Heisenberg and Josephson couplings. A theoretical analysis using a renormalized mean field theory, favors the former explanation. The potential for an improved determination of the antiferromagnetic and superconducting order parameters in an ideal single layer from zero field NMR and infrared Josephson plasma resonances in  multilayer cuprates is discussed.
 \end{abstract}

\maketitle

\section{Introduction}
\label{sec:introduction}

Although the cuprate superconductors have been studied for a quarter century, the exact form of the interplay of
antiferromagnetic(AF) and d-wave pairing(dSC) order at low hole doping is not known reliably. Experiments on very low
doped cuprates show insulating behavior and a critical hole concentration, $x_c$, for the onset of
superconductivity. The presence of a strongly varying potential due to the random distributions of the acceptors is
believed to be important at very low hole densities. There are reports of a spin glass phase separating the AF and dSC
regions of the phase diagram \cite{review}. A series of neutron scattering experiments on La$_{2-x}$Sr$_x$CuO$_4$ found
stripe order but with a change of orientation from parallel to the Cu-O-Cu bonds at $x > x_s$ to a $45^{\circ}$
direction at $x < x_s$ \cite{waki}. Neutron scattering experiments on YBa$_2$Cu$_3$O$_{6+y}$ found spin density wave
(SDW) order in the hole concentration range 0.05$< x < 0.08$\cite{haug}.  On the other hand coexistence of uniform AF
order with dSC has been deduced from NMR studies of multilayer Hg- and
Ba$_2$Ca$_3$Cu$_4$O$_8$(F$_y$O$_{1-y}$)$_2$-cuprates in a substantially larger hole density range\cite{kitaoka1} . In this paper we
analyse the role of AF and dSC interlayer coupling in multilayer cuprates and their effect on the coexistence on AF and
dSC order.

The observation by 
Kitaoka, Mukuda and collaborators \cite{kitaoka1} of substantial zero magnetic field shifts on a unique Cu-site in the innermost
layers led them to claim uniform AF order in underdoped layers with $x \sim 0.1$. The multilayer cuprates are
 doped by acceptors in the insulating blocks between the multilayer blocks,  but their random potential will be weakened in the  inner layers due to screening by the more highly doped metallic
outer layers. Thus these multilayer  cuprates offer the best possibility to reliably determine AF and dSC ordering at low doping in a clean single layer. In
this paper we illustrate the potential of multilayer systems by considering the case of  4-layer cuprates for which  experimental results exist in a wide density range.
  Variational Monte Carlo calculations for an ideal single layer strong coupling t-J model find a substantial
coexistence region of spatially uniform AF and dSC order up to a critical hole density $x \sim 0.1$
\cite{ogata}. Theoretical support for dSC order at low doping in a single layer also comes from exact diagonalization studies on clusters of the t-J
model containing up to 32-sites doped with 2 holes \cite{Leung}. When extrapolated to an infinite layer, these point to a robust d-wave cooperon ( bound hole pair ) resonance at
low doping . The virtual exchange of cooperons will act as a  pairing mechanism for doped holes in near nodal states at low hole densities, in the presence of strong AF local (and probably long range) order \cite{Rice}.

The two order parameters to be determined are the sublattice
magnetization of AF order and the pairing amplitude of dSC
ordering. The former is directly measured by the zero field shift
on Cu-sites  in NMR experiments. \cite{kitaoka1} In multilayer
samples the Josephson couplings between the dSC order in
neighboring layers lead to the Josephson plasma resonances which
appear in infrared spectra with the electric field oriented along
the c-axis \cite{Marel}. The energies of these resonances are
proportional to the product of the dSC order parameters. The
combination of these two measurements on the same samples can, in
principle, be used to determine the values and the dependence of
both AF and dSC order parameters on the hole density. To obtain
information on the interplay of AF and dSC order in a single
layer, we need to include the effects of interlayer couplings
between the respective order parameters in the analysis. Note,
the  interlayer spin coupling disfavors SDW order in adjacent
planes with different hole densities and therefore different
periodicities. As a result we restrict our attention to the case
of commensurate AF and dSC order.

   In this paper we report on calculations
using the renormalized mean field theory (RMFT),
a method introduced in the early days by Zhang and collaborators \cite{RMFT}, to treat the strong coupling t-J model. We extend the method to
 multilayer t-J models. RMFT calculations\cite{stripe-rvb} generally
 agree well with more accurate variational Monte-Carlo calculations \cite{VMC}
 in the case of a single layer.

   To treat multilayer materials we require the values
   of the interlayer couplings. The  Heisenberg spin-spin coupling
   has been directly measured with a strength typically
of order a tenth of the intralayer coupling. The interlayer Josephson coupling constants are less well known, as will be discussed further below.

 A second key input is the value of the hole densities
 in inequivalent layers. Kitaoka and collaborators have measured
 the Cu-Knight shift in each layer and used the
 room temperature value as input in an empirical formula
 to estimate the hole density in each layer.
 We shall return to these inputs below.

\section{$t-J$ Model for multilayer cuprates}
\label{sec:model}

  We start by defining the t-J model  for a single plane.
 $H_l^{t-J}$ for the $l^{th}$ layer is given by
\begin{align}
\label{eq:1}
H^{t-J}_l & = -\sum_{ij \sigma} P_G t_{ij} c_{l, i \sigma}^{\dag} c_{l, j \sigma} P_G + J\sum_{\left\langle ij
  \right\rangle} \mathbf{S}_{l, i} \cdot \mathbf{S}_{l, j},
\end{align}
where $c_{i l \sigma}$ is an annihilation operator of a spin
$\sigma$ electron at site $i$ in layer $l$, $P_G$ is the
Gutzwiller projection operator which enforces the no-double
occupancy condition in cuprates, $t_{ij}$ is the hopping integral
between site i and site j with nearest neighbor hopping $t_{NN} =
t$, next nearest neighbor hopping $t_{NNN} = t'$, and $t_{ij} = 0$
for the remaining site pairs.  J is the superexchange coupling
between NN sites.

    The interlayer coupling,
$H_{lm}$, between two neighboring layers $l$ and $m$ includes a
superexchange spin-spin coupling
 and an interlayer hopping term. The
interlayer superexchange coupling is given by
\begin{align}
\label{eq:2}
H_{lm}^J = J_{\perp} \sum_i \mathbf{S}_{l, i} \cdot \mathbf{S}_{m, i},
\end{align}
and the interlayer hopping term reads
\begin{align}
\label{eq:3}
H^{t}_{lm} & = P_G \sum_{\mathbf{k}} t_{\perp}(\mathbf{k}) (c_{l, \mathbf{k} \sigma}^{\dag} c_{m, \mathbf{k} \sigma} +
h.c.)  P_G,
\end{align}
where\cite{interlayer} $t_{\perp}(\mathbf{k}) = t_{\perp}\phi^2_{\mathbf{k}}/2$, with $\phi_{\mathbf{k}}=\cos k_x - \cos k_y$.

 In a system with more than two
layers, the hole concentrations on two adjacent inequivalent
layers are  different and the Fermi surfaces of the two layers are
mismatched. Because  planar momentum is conserved in interlayer
hopping, the mismatch of the Fermi surfaces means that direct interlayer
single particle hopping and interlayer pairing can be neglected.
However a  Cooper pair  can tunnel from the Fermi surface of one
layer into a pair state off the Fermi surface on a neighboring
layer, conserving the planar momentum of the individual electrons, This process can be written down in standard perturbation
theory as
\begin{align}
\label{eq:4}
H^{\Delta}_{lm} & = - \sum_{\mathbf{k}} \frac{t_{\perp}^2 \phi^4_{\mathbf{k}}}{2\omega_c}
\nonumber\\
& \phantom{=-} \times P_G \left(c_{l, -\mathbf{k}\downarrow}^{ \dag} c_{l, \mathbf{k}\uparrow}^{\dag}
  c_{m, \mathbf{k}\uparrow}c_{m, -\mathbf{k}\downarrow} + h.c. \right) P_G \tag{4'}
\end{align}
with $\omega_c$ a characteristic energy.  This excited state Cooper pair can subsequently relax to the Fermi
surface of the second layer through interactions such as the
J-term in the single plane Hamiltonian.  This virtual tunneling
process through pair states off the Fermi energy will be the
leading contribution of the interlayer Josephson coupling, arising
from Cooper pair hopping between inequivalent layers leading to an interlayer
Josephson coupling
\begin{align}
\label{eq:4p}
H^{\Delta}_{lm} & = \sum'_{\mathbf{k} \mathbf{k}'} \frac{t_{\perp}^2\phi^4_{\mathbf{k}}
  J(\mathbf{k}-\mathbf{k}')}{ 2 \omega_c \omega'_c}
\nonumber\\
& \phantom{=-} \times P_G \left(c_{l, -\mathbf{k}\downarrow}^{ \dag} c_{l, \mathbf{k}\uparrow}^{\dag}
  c_{m, \mathbf{k'}\uparrow}c_{m, -\mathbf{k'}\downarrow} + h.c. \right) P_G,
\end{align}
where $J(\mathbf{q}) = J (\cos q_x + \cos q_y)$ is the pair hopping amplitudes, $\omega'_c$ is a characteristic energy
related to the intralayer relaxation process, and the prime on the summation denote
s that the summation are only in the
vicinity of Fermi surfaces with cut-off $\omega_D$.

\begin{figure}[htbp]
\centerline{\includegraphics[width=0.26\textwidth]{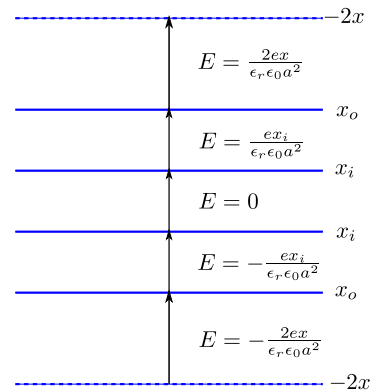}}
\caption[]{\label{fig:charge} Schematic illustration for the hole carrier charge distribution $x$, electric field
  $\vec E$ (direction indicated by arrows) in 4-layer Cu-oxides. Horizontal lines represent layers. Sub-indices $i$ and
  $o$ represent inner and outer planes, respectively. The top and bottom dashed lines represent chemically doped
  charge.}
\end{figure}

  In  multilayer materials  the distribution of hole densities on the individual layers plays an important role.
Theoretical estimates of the hole densities can be made using a Hartree approximation scheme by including the
electrostatic energy in the Hamiltonian \cite{electrostatic}. For simplicity, we consider each layer as an infinite
plane with homogeneous charge. The electric field generated by such a plane is $E = \frac{\rho}{2 \epsilon_r
  \epsilon_0} = \frac{e x }{2 \epsilon_r \epsilon_0 a^2}$, where $\epsilon_r$ is the relevant dielectric constant,
$x$ is the hole concentration of the layer, and $a$ is the lattice constant.  For a 4-layer system, we denote
$x_o$ and $x_i$ as the hole concentrations of the outer layers and the inner layers, respectively.  The
average hole concentration is $x = (x_o + x_i) / 2$, while the electron concentration of the acceptor
layer above and below the 4-layers is $2 x$.  The spatial distribution of electric field is depicted in fig.~\ref{fig:charge}.  We are only
interested in the terms relevant to the charge inhomogeneity, so the electrostatic energy between acceptor layer and outer
layer can be ignored.  Then the electrostatic energy per plaquette in the 4-layer case is $E_{es} = \frac{ e^2
  x_i^2 d}{\epsilon_r \epsilon_0 a^2} $, where $d$ is the distance between two adjacent CuO$_2$ layers.  The Hamiltonian for multilayer system then reads
\begin{align}
\label{eq:5}
H & = \sum_l H_l^{t-J} + \sum_{\left\langle lm \right\rangle} (H^{\Delta}_{lm} + H^J_{lm}) + E_{es}.
\end{align}
   The dielectric screening of the interlayer electric field arises from displacements of the ionic layer between the neighboring layers and is not identical to the static c-axis dielectric screening due to optical phonons etc. We adjust the value of $\epsilon_r$ to fit the estimated values for the layer densities (see below).

\section{Renormalized mean field theory for multilayer system}
\label{sec:rmft-mult-syst}

In this section, we will use RMFT to analyse the multi-layer Hamiltonian. The main element of the RMFT theory is the
so-called Gutzwiller approximation, which replaces the Gutzwiller projection in the Hamiltonian by simple numerical
factors. Let $\left\langle \hat{O} \right\rangle $ and $\left\langle \hat{O} \right\rangle_0$ be the average values of
operator $\hat{O}$ in the Gutzwiller projected state and in the unprojected state, respectively, then $\left\langle
  \hat{O} \right\rangle \approx g_o \left\langle \hat{O} \right\rangle_0$, with $g_o$ a numerical factor, depending on
the hole density and the process associated with the operator $\hat{O}$. The coexistence of AF and dSC order in the
t-t'-J model for a single layer material has been investigated within RMFT recently by K.-Y. Yang et
al. \cite{stripe-rvb}.  Here we generalize their calculations to a multilayer system. Note that in the RMFT the
superconducting order parameter is non-zero at wave-vector off the Fermi surface. The virtual tunneling process through
pair states off the Fermi energy when the two layers have different hole concentrations we discussed in Section II may
also be taken into account by including the pair scattering on the same layer along with the second order interlayer
hopping process conserving momentum of single electrons. Below we shall use Eq. (4') for the interlayer tunneling
directly to study the multi-layer system described in Eq. (5).

We now introduce the following mean fields
\begin{align}
\label{eq:7}
\Delta_{0, l} &= \frac{1}{4 N} \sum_i \left\langle \sum_{\delta = \pm \hat{x}}c_{l, i \uparrow} c_{l, i + \delta
    \downarrow} - \sum_{\delta = \pm \hat{y}} c_{l, i \uparrow} c_{l, i + \delta \downarrow}\right\rangle_0
\nonumber\\
m_{0, l} &= \frac{1}{N} \sum_i (-1)^i \left\langle S_{l, i}^z \right\rangle= \frac{1}{2 N} \sum_{i \sigma} \sigma (-1)^i
\left\langle c_{l, i \sigma}^{\dag} c_{l, i \sigma} \right\rangle_0
\nonumber\\
\chi_{0, l, ij, \sigma} &= \left\langle c_{l, i \sigma}^{\dag} c_{l, j \sigma} \right\rangle_0,
\end{align}
where $\left\langle \right\rangle_0$ indicates average with the unprojected state.  Since we are only interested in the
ground state, we can assume that the mean fields $\Delta_{0, l}$ are all real.  In the presence of AF order, the
magnetization on the sublattices A and B are opposite, and the mean fields $\chi_{0, l, ij, \sigma}$ from NNN sites
depend on the sublattice and spin.  We introduce the following mean fields, $\chi_{0, l, AA, \sigma} = \frac{1}{2N}
\sum_{i \in A} \sum_{j = NNN(i)} \chi_{0, l, ij, \sigma} $ and $\chi_{0, l, BB, \sigma} = \frac{1}{2N} \sum_{i \in B}
\sum_{j = NNN(i)} \chi_{0, l, ij, \sigma} $. Since $\chi_{0, l, AA, \sigma} = \chi_{0, l, BB, -\sigma}$, we can further
simplify the notation as $\chi_{0, l, AA} = \chi_{0, l, AA, \uparrow}$ and $\chi_{0, l, BB} = \chi_{0, l, BB,
  \uparrow}$.  On the other hand, the mean fields $\chi_{0, l, ij, \sigma}$ for NN sites are independent of the
sublattice $ij$ and spin $\sigma$, so we can set the mean fields $\chi_{0, l, AB} = \frac{1}{8N} \sum_{i \sigma} \sum_{j
  = NN(i)} \chi_{0, l, ij, \sigma}$.  The hole concentration on each layer satisfies the condition.
\begin{align}
\label{eq:8}
x_l & = 1 - \left\langle n_l \right\rangle = 1 - \frac{1}{N} \sum_{i \sigma} \left\langle c_{l, i \sigma}^{\dag}
  c_{l, i \sigma} \right\rangle.
\end{align}

Then we consider the Gutzwiller factors.  In intralayer terms, we adopt the same form of the factors $g_{l, AB}^t, g_{l,
  AA}^t, g_{l, BB}^t, g_l^m, g_l^{xy}$, and $g_l^z$ as used previously for the single layer t-J model \cite{stripe-rvb}.
For the interlayer hopping and superexchange coupling terms, we make a simple assumption that the Gutzwiller factors are
$g_{l m}^t = \sqrt{g^t_{l, AB} g^t_{m, AB}}$ and $g^z_{l m} = \sqrt{g^m_l g^m_m}$, respectively.  The total energy reads
\begin{align}
\label{eq:9}
E &= \sum_l E_l + \sum_{\left\langle lm \right\rangle} E_{lm} + E_{es},
\end{align}
where $E_l$ is given by the RMFT for single layer t-t'-J model \cite{stripe-rvb}, the electrostatic energy $E_{es}$ is
specified in the previous section, and the interlayer energy $E_{lm}$ takes the form
\begin{align}
\label{eq:10}
E_{l m} &= \frac{\left\langle H_{l m}^{\Delta} + H^J_{l m} \right\rangle}{N} = -4 (g^t_{l m})^2\frac{t_{\perp}^2}{4 N
  \omega_c} \sum_{\mathbf{k}}' \phi_{\mathbf{k}}^4 \bigl( \Delta_{0, l, \mathbf{k}} \Delta_{0, m, \mathbf{k}}
\nonumber\\
& \phantom{=}+ \Delta_{0, l, \mathbf{k+Q}} \Delta_{0, m, \mathbf{k+Q}} \bigr) + J_{\perp} g^z_{l m} m_{0, l}
\bar{m}_m,
\end{align}
with $\Delta_{0, l, \mathbf{k}} = \left\langle c_{\mathbf{k}\uparrow}^l
  c_{-\mathbf{k} \downarrow}^l \right\rangle_0$.  Note that the magnetic moment $m_l$ and superconducting
order parameter $\Delta_l$ are related to the mean fields by the Gutzwiller renormalization factors\cite{RMFT},
\begin{align}
\label{eq:11}
m_l & = \sqrt{g^z_l} m_{0, l} & \Delta_l & = g^t_{l, AB} \Delta_{0, l}.
\end{align}

These approximations lead to the mean field Hamiltonian
\begin{align}
\label{eq:12}
H_{MF} &= \sum_{l \mathbf{k} \sigma} \left(\epsilon_{l, \mathbf{k}}c_{l,\mathbf{k} \sigma}^{\dag} c_{l, \mathbf{k}
    \sigma} - \sigma M_{l, \mathbf{k}} c_{l \mathbf{k} \sigma}^{\dag} c_{l, \mathbf{k+Q}
    \sigma}  \right) \nonumber\\
& \phantom{=} - \sum_{l \mathbf{k} \sigma}\left( V_{l, \mathbf{k}} c_{l, -\mathbf{k} \downarrow} c_{l, \mathbf{k}
    \uparrow} + h.c. \right),
\end{align}
where
\begin{align}
\label{eq:13}
\epsilon_{l, \mathbf{k}} &= \frac{\partial E}{4 \chi_{0, l, AB}} \gamma_{\mathbf{k}} + \frac{1}{2} \left(
  \frac{\partial E}{\partial \chi_{0, l, AA}} + \frac{\partial E}{\partial \chi_{0, l, BB}} \right)
\theta_{\mathbf{k}} - \tilde{\mu}_l
\nonumber\\
M_{l, \mathbf{k}} & = - \frac{1}{2} \left( \frac{\partial E}{\partial \chi_{0, l, AA}} - \frac{\partial E}{\partial
    \chi_{0, l, BB}} \right) \theta_{\mathbf{k}} - \frac{1}{2} \frac{\partial E }{\partial m_{0, l}}
\nonumber\\
V_{l, \mathbf{k}} &= \frac{1}{4} \frac{\partial E_l}{\partial \Delta_{0, l}} \phi_{\mathbf{k}} + \frac{N}{2} \sum_m
\frac{\partial E_{lm}}{\partial \Delta_{0, l, \mathbf{k}}}
\nonumber\\
\tilde{\mu}_l &= \frac{\partial E}{\partial x_l} + \mu,
\end{align}
and $\gamma_{\mathbf{k}} = \cos k_x + \cos k_y$, and
$\theta_{\mathbf{k}} = \cos k_x \cos k_y$.

\section{Results and Discussions}
\label{sec:results-discussions}

\begin{figure}[htbp]
\centerline{\includegraphics[width=0.4\textwidth]{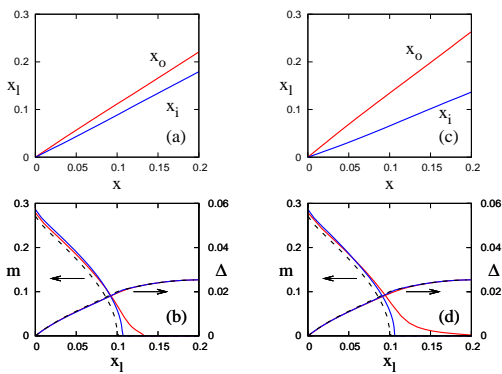}}
\caption[]{\label{fig:orders} The order parameters and hole concentrations calculated in 4-layer t-t'-J model Eq. (5).
  Model parameters are $J=0.3 t$, $t' = -0.3 t$, interlayer pair hopping $t_{\perp}^2/\omega_c = 0.1 t$, interlayer spin
  coupling $J_{\perp} = 0.1 J$.  Upper panel: the hole concentrations for each layer $x_l$ vs average hole
  concentrations $x$.  Thick blue curves are for inner layer ($x_i$), thin red for outer layer ($x_o$).
  Lower panel: magnetic moment $m$ and superconducting order parameter $\Delta$ as function of hole doping for each
  layers $x_l$.  Thick blue curves are for inner layer, thin red for outer layer.  Left panel: dielectric constant
  $\epsilon =200$, right panel: $\epsilon =50$.  Dashed curves in (b) and (d) are for single layer results, plotted for
  comparison.}
\end{figure}

  The mean field energy defined by Eqn (8) has been minimized
  numerically for the case of 4-layers.
  The results depend on the hole densities which in
  turn depend on the value of the dielectric constant,
$\epsilon_r$.  In fig.~\ref{fig:orders}, the two upper figures
show the density in each layer as function of the total doping for
two choices $\epsilon_r = 200$ and $\epsilon_r = 50$. The lower
figures show the resulting values of the order parameters in each
layer. The single layer result is also shown in the figure as a
dashed line for comparison. The SC order parameters of the 4-layer
system and single layer system are almost equal and independent of
$\epsilon_r$.  This is a consequence of the weak Josephson
coupling between adjacent layers in the multilayer system due to
the mismatch of the Fermi surfaces and the suppression of the
interlayer hopping by the strong onsite Coulomb interaction.

\begin{table}[htbp]
  \caption[]{\label{tab:epsilon} Comparison between NMR experimental results
    on Ba$_2$Ca$_3$Cu$_4$O$_8$(F$_y$O$_{1-y}$)$_2$\cite{kitaoka1} and the renormalized mean field
    theory for 4-layer $t-t'-J$ model. $y$ is the content of $F$ in
    the compound. $x_{op}$ and $x_{ip}$ are the hole
    concentrations on the outer and inner planes. $m_{op}$ and
    $m_{ip}$ are the AF magnetic moment on the outer and inner planes,
    respectively. $\epsilon$ is the fitted value of the dielectric
    constant from the hole distribution. $\omega_{ii}$ and $\omega_{io}$ are the Josephson plasma frequencies between two inner
    layers and between inner layer and outer layer, respectively.} \vspace{4mm}
\begin{tabular}{|c|c|c|c|c||c|c|c|c|c|}
  \hline \multirow{2}{*}{y} & \multicolumn{4}{c||}{NMR Exp.} & \multicolumn{5}{c|}{Theory} \\
  \cline{2-5}  \cline{6-10} & $x_{op}$ & $m_{op}$ & $x_{ip}$ & $m_{ip}$ &
  $\epsilon$ & $m_{op}$& $m_{ip}$  & $\omega_{io}$ & $ \omega_{ii}$\\
  \hline 0.6 & 0.141 & 0 & 0.089 & 0 & 87 & 0.010 & 0.096 & 492 & 394 \\
  \hline 0.7 & 0.111 & 0 & 0.074 & 0.04 & 107 & 0.043 & 0.137 & 399 & 322 \\
  \hline 0.8 & 0.092 & 0 & 0.069 & 0.06 & 170 & 0.093 & 0.150 & 348 & 301\\
  \hline 1 & 0.073 & 0.055 & 0.059 & 0.09 & 247 & 0.138 & 0.173 & 278 & 243\\\hline
\end{tabular}
\end{table}

 The interlayer AF coupling slightly enhances the AFM moment and critical doping $x_c$ of the inner layers and leads to a finite AFM moment on the outer layers when their hole
concentration is much larger than the single layer critical doping $x_c$.  Such a long tail arises
because of the lower hole concentration on the inner layers. A  smaller $\epsilon_r$ leads to larger charge imbalance between the outer and inner layers and so to larger $x_c$ of outer layers, as
shown in fig.~\ref{fig:orders}.  If one neglects the long tail of the AFM moment of the outer layers, the phase diagram
of multi-layer system is very similar with that of a single layer system, independent of the value of the dielectric
constant $\epsilon_r$.  This indicates that  experiments on multilayer system can capture the essential physics of
single layer t-J model

 We turn now to  the relation of the model with  experiment.  The two order parameters we investigated above are the
AFM magnetization and the pairing amplitude of dSC ordering.  The former has been directly  measured by the zero field shift
in NMR.  In the NMR experiments, Shimizu, Kitaoka and collaborators \cite{kitaoka2} have also used an empirical scaling of the Cu
Knight shift, measured at room temperature, to deduce hole densities of individual layers. Note this scaling form has evolved with time.  In tab.~\ref{tab:epsilon}, we
show the value of the relative dielectric needed to fit the charge distribution recently determined from NMR experiments on the 4-layer
Ba$_2$Ca$_3$Cu$_4$O$_8$(F$_y$O$_{1-y}$)$_2$ cuprate.  Note, in these oxy-fluoride cuprates the hole density differs substantially from the nominal  F- concentration,y. Though the dielectric constant required to fit the hole density changes from 87 to 247 as the total hole density decreases from 0.46 to 0.264,  the physics in each layer should not change very much as analyzed above.  In Table 1, we also
show the AFM moments calculated by the RMFT and measured by NMR experiments.  The former values are considerably larger
than the latter, indicating an over estimation of the AFM order in the RMFT theory.  The comparison between theory and
experiment is sensitive to the values of the hole density. For example, if we scale the value of the estimated hole
density so that the optimal density with the highest Tc for a single layer is $( x_c = 0.2)$  rather  $(x_c = 0.16)$
quoted in Mukuda et al.\cite{kitaoka1,kitaoka2}, this leads to an increase in the input values for the hole density in the RMFT calculations by a factor of 1.25. Then the calculated values of the AF moment shown in Tab 1, are smaller, although still too big, and in better agreement with experiment.

  Turning to the dSC ordering, as mentioned earlier the pairing amplitudes enter into  the Josephson plasma frequencies observed in the
infrared spectra with the electric field oriented along the
c-axis.  In multilayer cuprates, there are two kinds of Josephson
plasma frequencies which correspond to inter-multilayer and
intra-multilayer Josephson couplings respectively.  In this paper,
we are only interested the latter, i.e. the Josephson plasma
frequencies related to the Josephson coupling between two adjacent
CuO$_2$ layers in same unit cell.  The Josephson energy between
those two layers can be written as $E_J = -J_{lm} \cos(\phi_l -
\phi_m)$, where $\phi_l$ and $\phi_m$ are the phase of the
superconducting order parameter of the two layers respectively.
For the two layers with the same hole concentration, $J_{lm}$ is
given in our RMFT by,
\begin{align}
\label{eq:14}
J_{lm} & = \frac{2}{\beta N} \sum'_{\mathbf{k}}\sum_{i \omega_n}
(g^t_{lm})^2 t^2_{\perp} (\mathbf{k})
\Bigl|\mathcal{F}_l(\mathbf{k}, i \omega_n)
\Bigl|\mathcal{F}^{\dag}_m(\mathbf{k}, i \omega_n) \nonumber\\
& \phantom{\propto} + \Bigl|\mathcal{F}_l(\mathbf{k+Q}, i
\omega_n) \Bigl|\mathcal{F}^{\dag}_m(\mathbf{k+Q}, i
\omega_n)\Bigr| = t_{\perp}^2 \tilde{\omega}^2_{lm},
\end{align}
where $\mathcal{F}_l$ is the anomalous Green's function of layer $l$, $i \omega_n$ is the Matsubara frequency,
$\mathbf{Q} = (\pi, \pi)$, and $\sum'$ indicates the summation is only over the reduced zone.
The Josephson plasma frequency which corresponds to the Josephson coupling between those two layers is proportional to
the Josephson coupling constant \cite{Leggett,Marel}, i.e. $\omega^J_{lm} \propto \sqrt{J_{lm}}$.  So we have
\begin{align}
\label{eq:15}
\omega^J_{lm} &=  C \tilde{\omega}_{lm},
\end{align}
where $C$ is a constant related to $t_{\perp}$ and finite frequency dielectric constant $\epsilon(\omega_J)$.  For
simplicity, we can assume that $C$ is weakly dependent on the hole concentrations and materials.

For a 4-layer material, there are two intra-unit cell Josephson plasmon frequencies $\omega_{io}$ and $\omega_{ii}$
which correspond to the Josephson coupling between two inner layers and between the inner and outer layers respectively.
A recent optical experiment on a 4-layer Hg-based compound has shown that the Josephson plasmon frequencies are 360
cm$^{-1}$ for inner layers and 540 cm$^{-1}$ for outer and inner layers respectively leading to a ratio of
$1.5$\cite{uchida}.  According to eqn.~\eqref{eq:15}, $\omega_{io}/\omega_{ii} =
\tilde{\omega}_{io}/\tilde{\omega}_{ii}$.  Using the estimated hole concentrations of the Hg-compound, the ratio
calculated with RMFT is around $1.25$ which is fairly close to the experimental value.  To make further comparison, we
calculate the factor $C$ with $C = \sqrt{\omega_{io} \omega_{ii}/\tilde{\omega}_{io} \tilde{\omega}_{ii}} \approx 10$,
where the plasmon frequencies are the experimental values for the Hg-compound and the $\tilde{\omega}$ is calculated with
RMFT.  Using this estimate for $C$ allows us to obtain values for the plasma frequencies for the 4-layer Ba$_2$Ca$_3$Cu$_4$O$_8$(F$_y$O$_{1-y}$)$_2$ material.
The results are quoted in Table \ref{tab:epsilon}.

\section{Summary}
\label{sec:summary}

   In this paper we presented a series of calculations using the RMFT to treat the AF and dSC order parameters in 4-layer cuprates. The RMFT calculations/cite{stripe-rvb} and also the results of single VMC calculations/cite{VMC}, give a more extended region of coexistence of the two order parameters than that observed in single and bilayer cuprates. The extended coexistence region in the 4- layer cuprates agrees with the conclusion of the NMR experiments in the inner layers. In our analysis the extended region comes not from strong interlayer magnetic and Josephson coupling,  but from the extended coexistence region already present in single layers in the RMFT approximation. This analysis suggests that the difference in the density range of coexisting order in multilayer cuprates compared to experiments on single and bilayer cuprates, is due to the better screening of the external potential modulation  in the inner layers of multilayer cuprates, when compared to  single and bilayer cuprates. The latter are adjacent to the random acceptors, while the metallic outer layers screen the random potential at the inner layers.
  In principle the combination of zero field NMR and infrared Josephson plasma frequencies allows one to determine both AF and dSC order parameters. However, in order to go further and directly compare measured and calculated order parameters a better determination of the interlayer Josephson coupling constant and its variation with the hole densities in the individual layers, is required. To this end a series of measurements of the Josephson plasma frequencies and the AF moments by zero field NMR at different hole densities on the same multilayer material would be helpful.

We acknowledge financial support in part from Hong Kong RGC GRF grant HKU706507, the National Natural Science
Foundation of China 10804125, and also from the Swiss Nationalfond and MANEP network.


\begin{thebibliography}{99}
\bibitem{review} For a review see M.-H.Julien,  Physica B \textbf{329-333}, 693 (2003)
\bibitem{waki} S. Wakimoto, R. J. Birgeneau, M. A. Kastner, Y. S. Lee, R. Erwin, P. M. Gehring, S. H. Lee, M. Fujita,
  K. Yamada, Y. Endoh, K. Hirota, and G. Shirane, Phys. Rev. B \textbf{61}, 3699 (2000)
\bibitem{haug} D. Haug, V. Hinkov, Y. Sidis, P. Bourges, N. B. Christensen, A. Ivanov, T. Keller, C. T. Lin, and B. Keimer, New
  J. Phys \textbf{12}, 105006 (2010)
\bibitem{ogata} M. Ogata and H.Fukuyama, Rep. Prog. Phys. \textbf{71}, 036501 (2008)
\bibitem{Leung} A. L. Cheryshev, P. W. Leung, and R. J. Gooding, Phys. Rev. B \textbf{58}, 13594 (1998)
\bibitem{kitaoka1} For a recent review see H. Mukuda, S. Shimizu, A. Iyo, and Y. Kitaoka, J. Phys. Soc. Jpn. \textbf{81}, 011008 (2012)
\bibitem{Leggett} A. J. Leggett, Prog. Theor. Phys. \textbf{36}, 901-930 (1966)
\bibitem{Marel} D. van der Marel, J. Supercond. \textbf{17}, 559 (2004)
\bibitem{RMFT} F. C. Zhang, C. Gros, T. M. Rice, and H. Shiba, Supercond.Sci.Technol. \textbf{1}, 36 (1988)
\bibitem{stripe-rvb} K.-Y.Yang, W. Q. Chen, T. M. Rice, M. Sigrist, and F. C. Zhang, New J. Phys. \textbf{11},
  055053 (2009)
\bibitem{VMC} G. J. Chen, R. Joynt, F. C. Zhang, and G. Gros, Phys. Rev. B \textbf{42}, 2662 (1990);T. Giamarchi and
  C. Lhuillier, Phys. Rev. B \textbf{43}, 12943 (1991) ; A. Himeda and M. Ogata, Phys. Rev. B \textbf{60}, R9935 (1999)
\bibitem{interlayer} S. Chakravarty, A. Sudbo, P. W. Anderson, and S. Strong, Science \textbf{261}, 337 (1993);
  O. K. Andersen, A. I. Liechtenstein, O. Jepsen, and F. Paulsen, J. Phys. Chem. Solids \textbf{56},1573, (1995)
\bibitem{electrostatic} M. Di Stasio, K. A. Mueller, and L. Pietronero, Phys. Rev. Lett \textbf{64}, 2827 (1990)
\bibitem{Gan} J. Y. Gan, Y. Chen, Z. B. Su, and F. C. Zhang, Phys. Rev. Lett. \textbf{94}, 067005 (2005)
\bibitem{ogatafactor} M. Ogata and A. Himeda, J. Phys. Soc. Jpn. \textbf{72}, 374 (2003)
\bibitem{kitaoka2} S. Shimizu, S. Iwai, S. Tabata, H. Mukuda, Y. Kitaoka, P. M. Shirage, H. Kito, and A. Iyo,
  Phys. Rev. B \textbf{83}, 144523 (2011)
\bibitem{uchida} Y. Hirata, K. M. Kojima, S. Uchida, M. Ishikado, A. Iyo, H. Eisaki,
  and S. Tajima, Physica C: Superconductivity 470, S44 (2010)
\bibitem{Rice} T. M. Rice, K.-Y. Yang, and F. C. Zhang, Rep. Prog. Phys. \textbf{75}, 016502 (2012)

\end{thebibliography}
\end{document}